\documentstyle[mprocl,psfig]{article}

\def\Journal#1#2#3#4{{#1} {\bf #2}, #3 (#4)}

\def\PRL{\em Phys. Rev. Lett.}
\def\PRD{{\em Phys. Rev.} D}

\def\etal{\it et al.}
\def\be{\begin{equation}}
\def\ee{\end{equation}}

\begin{document}

\title{THE `UPS' AND `DOWNS' OF A SPINNING BLACK HOLE}

\author{C.M. CHAMBERS, W.A. HISCOCK, B.E. TAYLOR}

\address{Department of Physics, Montana State University, 
         Bozeman, MT 59717-3840, USA}

\maketitle

\abstracts{We report and comment upon the principal results 
of an investigation into the evolution of rotating black holes 
emitting massless scalar radiation via the Hawking process. It 
is demonstrated that a Kerr black hole evaporating by the
emission of scalar radiation will evolve towards a state with 
$a \approx 0.555M$. If the initial specific angular momentum
is larger than this value the hole will spin down to this value;
if it is less it will spin up to this value.
The addition of higher spin fields to the picture strongly 
suggests the final asymptotic state of a realistic evaporation 
process will be characterized by an $a/M = 0$.
}

%
%

\section{Introduction}
It is generally accepted that an isolated black hole
evaporating via the Hawking radiation mechanism will,
regardless of its initial state, approach a final asymptotic 
state described by the Schwarzschild solution. Indeed, the 
evolution of rotating black holes radiating fields of 
spin $s$ has been studied in some detail by 
Page.\cite{cmc-b10} For a Kerr black hole,
emitting a fixed set of fields with $s > 0$, Page has
shown that the angular momentum loss proceeds more quickly
than the mass loss, resulting in a final evolutionary state with
$a_{*} \equiv a/M = 0$, which is indeed characterized by the 
Schwarzschild solution. However, Page's work does suggest that 
a black hole emitting scalar fields $( s = 0 )$ might evolve differently
and approach a state with $a_{*} > 0$. For $s > 0$, it was found
that the dominant emission modes at low $a_{*}$ were those
with $\ell = m = s$.\footnote{Here $\ell$ and $m$ are 
the usual angular eigenvalues associated with
orbital angular momentum.} 
If the same holds true for
scalar fields then the dominant mode will be $\ell = 0$,
which carries off energy but no angular momentum. In that
case, one can envisage a situation in which the mass loss
occurs much more rapidly than the angular momentum loss and
leads to a final state with $a_{*}$ nonzero.

Here, we summarize the main results and conclusions of a
numerical study into the evolutionary path of a rotating black hole
emitting massless scalar fields via the Hawking process.\cite{cmc-b20}
For a Kerr black hole emitting only scalar radiation, we find 
the rate at which mass and angular momentum are lost leads
to a final evolutionary state characterized by an $a_{*} 
\approx 0.555$. Further to this we determine the number of
scalar fields required to cause a black hole, emitting 
a fixed set of higher spin fields, to evolve to a state
of nonzero $a_{*}$.

\section{Results}
The evolution of an isolated Kerr black hole, evaporating via
the Hawking mechanism, is represented by the rates at which its mass 
$(M)$ and angular momentum $(J)$ decrease. Following Page we 
express these rates in terms of the scale invariant quantities 
$f(a_{*}) \equiv -M^2 dM/dt$ and $g(a_{*}) \equiv -M a^{-1}_* dJ/dt$, 
where $t$ is the usual Boyer-Lindquist time coordinate for the Kerr 
spacetime.\footnote{Expressions for $f$ and $g$, and
the numerical procedures used to calculate them, can be
found in the work of Taylor \etal~\cite{cmc-b30}}
%
%
\begin{figure}[ht]
\centerline{
\psfig{figure=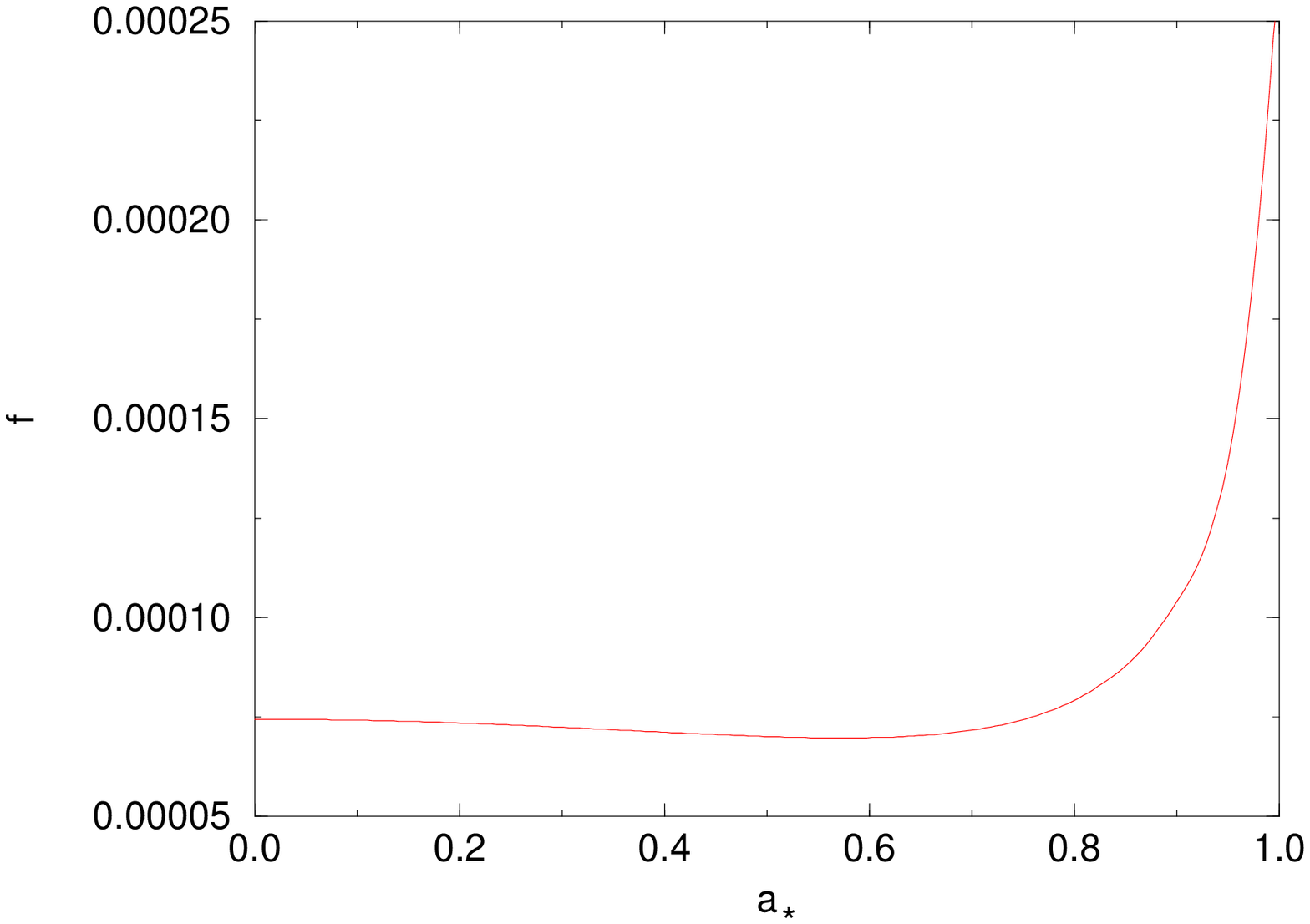,height=1.9in} 
\psfig{figure=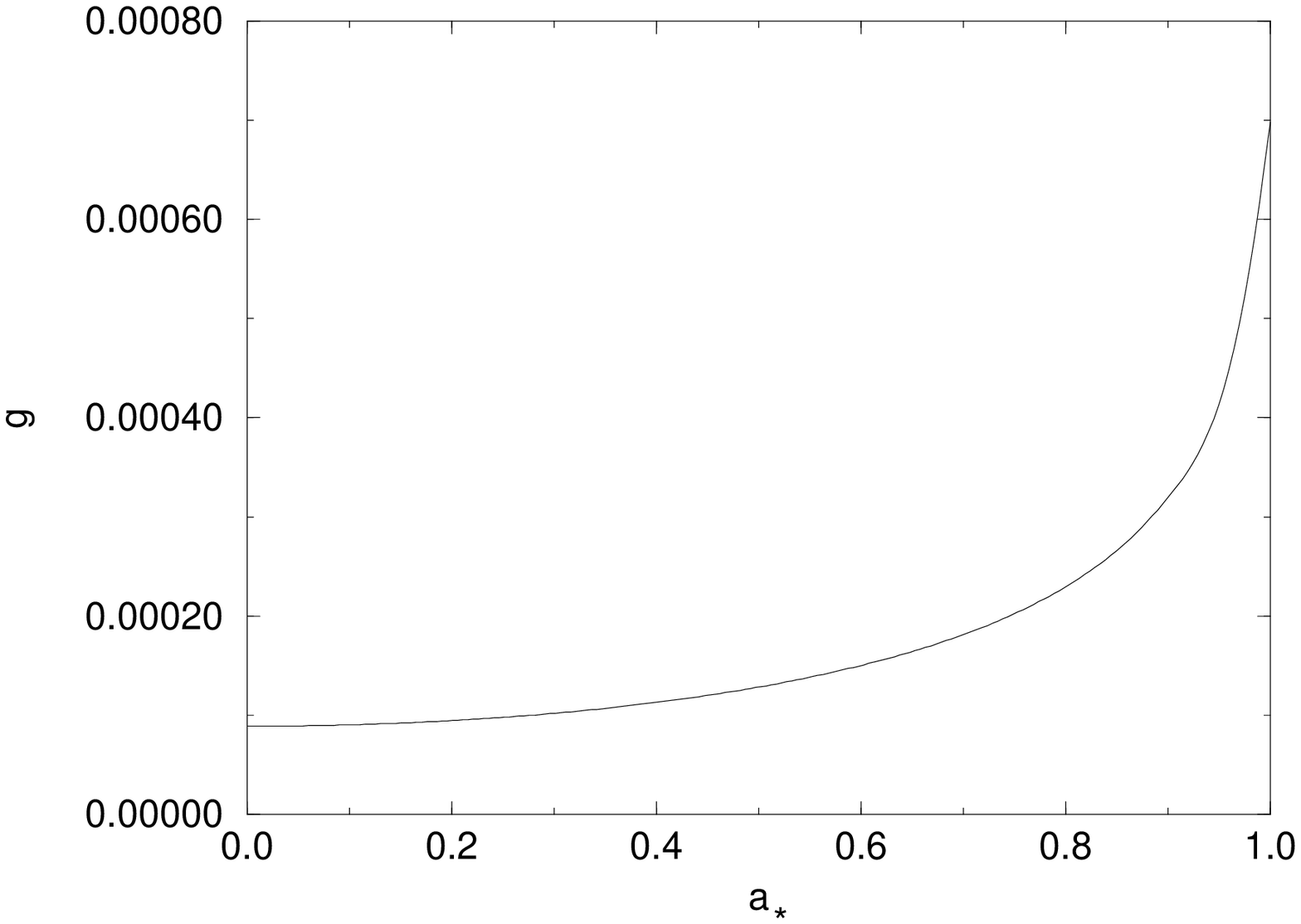,height=1.9in}}
\caption{The scale invariant mass and angular momentum loss
rates, $f$ and $g$, versus $a_{*}$ due to a single scalar 
field. The behavior of these functions is explained in the 
text.
\label{cmc-f10}}
\end{figure}
Figure~\ref{cmc-f10} shows the scale invariant mass loss rate 
$f$ and the scale invariant angular momentum loss rate $g$ of 
a Kerr black hole evaporating via purely scalar radiation. Unlike the 
higher spin case,\cite{cmc-b10} where both $f$ and $g$ 
are monotonically increasing functions of $a_{*}$, the mass
loss rate due to the emission of massless scalar radiation alone,
can clearly be seen to possess a minimum at $a_{*} \approx 0.574$.
This exceptional behavior can be ascribed to the fact that a
scalar field, unlike higher spin fields, is able to radiate in
an $\ell =0$ mode, which is not superradiant.  Our numerical 
study shows that the mass loss rate, due solely to the $\ell = 0$
mode, is a monotonically decreasing function of $a_{*}$ and that
at low $a_{*}$ this mode dominates the mass loss experienced
by the black hole. As $a_{*}$, and hence superradiance, increases 
the contribution of higher $\ell$ modes to the mass loss 
becomes significant and dominates at high $a_{*}$. The combined effect at 
intermediate $a_{*}$ is the observed minimum. The angular momentum 
loss rate $g$ can be seen to be a monotonically increasing function of
$a_{*}$. Since the $\ell = 0$ mode cannot carry off angular
momentum, the behavior of $g$ can be understood purely in
terms of superradiance. As $a_{*}$ and the effects of
superradiance increase more angular momentum is efficiently 
extracted from the hole. 
%
%
\begin{figure}[ht]
\centerline{
\psfig{figure=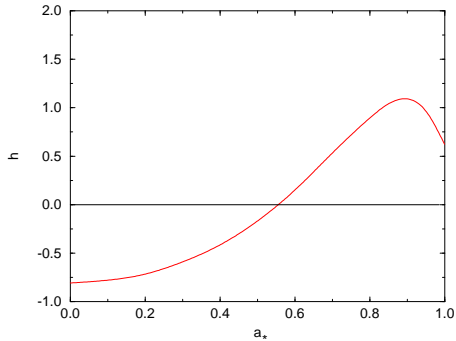,height=2.0in}}
\caption{The quantity $h$ is plotted versus $a_*$.  The zero of $h$
indicates that a black hole emitting purely massless scalar radiation will
evolve towards a state described by a value of $a_{*} \approx 0.555$.
\label{cmc-f20}}
\end{figure}

In order to gauge the relative loss rate of mass to angular
momentum, we define the scale invariant quantity, $h(a_{*})$, 
by
	\be 
          h(a_*) \equiv \frac{d \ln a_{*}}{d \ln M} = 
	  \frac{g(a_*)}{f(a_*)} - 2 \; .  
	\label{cmc-e10} 
	\ee
Figure~\ref{cmc-f20} shows $h(a_{*})$ versus $a_{*}$ for a Kerr black hole
emitting purely scalar radiation. The important feature of this plot
is the existence of a zero in $h$ at a finite, but nonzero, value of
$a_{*}$, namely $a_{*}\approx 0.555$. The rate of change of $a_{*}$
with respect to time is given by
	\be
	  \frac{da_{*}}{dt} = - \frac{a_{*} h f}{M^3}  \; .
        \label{cmc-e20} 
        \ee 
Since the black hole is in isolation it can gain neither mass nor
angular momentum and so $f$ and $g$, by definition, must be
positive definite functions of $a_{*}$ -- as demonstrated in 
Figure~\ref{cmc-f10}. A black hole with $0.555 < a_{*} < 1$ 
will have a positive value of $h$ and thus Eq.~\ref{cmc-f20} 
implies that $a_{*}$ will decrease over time until it reaches 
an asymptotic state characterized by an $a_{*} \approx 0.555$. 
Similarly, a black hole with $0 < a_{*} < 0.555$ has a negative 
value of $h$ and so $a_{*}$ will increase until the black hole 
achieves a state with $a_{*} \approx 0.555$.  Therefore, the 
existence of a zero in $h$ at $a_{*} = 0.555$ implies that 
a Kerr black hole will evolve to and remain at this value of $a_{*}$.

It is interesting to ask to what extent the inclusion of 
scalar field emission in the evaporation processes studied
by Page affects the evolution. Utilizing the results
for higher spin fields,\cite{cmc-b10} we have calculated 
the number of scalar fields in conjunction with a fixed
set of higher spin fields that results in a final state
with $a_{*} > 0$. For the set consisting of 1 spin-2,
1 spin-1 and 3 spin-1/2 fields, we find that $32$ scalar
fields are required. As no massless scalar fields have
yet been discovered in Nature it appears that
the conventional view of black hole evaporation will
continue to hold in any realistic evaporation process.

\section*{Acknowledgments}
CMC is a fellow of The Royal Commission For The Exhibition Of 1851 and 
gratefully acknowledges their financial support.  Travel to Israel for 
CMC was supported by NSF Grant Nos. PHY-9722529 to North Carolina State
University, and PHY-9511794 to Montana State University. 
The work of WAH and BT was supported in part by NSF Grant No. PHY-9511794.

\end{document}